\documentclass[12pt]{article}
\usepackage{graphicx}
\usepackage{amsmath}
\usepackage{amssymb}
\usepackage{caption2}
\begin{document}
\bibliographystyle{prsty}
\begin{center}
{\large {\bf \sc{   Structure of the axial-vector meson
$D_{s1}(2460)$ and the strong coupling constant $g_{D_{s1} D^* K}$
with the light-cone QCD sum rules }}} \\[2mm]
Z. G. Wang \footnote{E-mail,wangzgyiti@yahoo.com.cn.  }    \\
Department of Physics, North China Electric Power University,
Baoding 071003, P. R. China
\end{center}

\begin{abstract}
In this article, we take the point of view that the  charmed
axial-vector meson $D_{s1}(2460)$ is the conventional $c\bar{s}$
meson and  calculate the strong coupling constant $g_{D_{s1} D^* K}$
in the framework of the light-cone QCD sum rules approach.  The
numerical values  of strong coupling constants $g_{D_{s1} D^* K}$
and  $g_{D_{s0} D K}$ are very large, and support the hadronic
dressing mechanism. Just like the scalar mesons $f_0(980)$ and
$a_0(980)$, the scalar meson $D_{s0}(2317)$ and axial-vector meson
$D_{s1}(2460)$ may have small $c\bar{s}$ kernels  of the typical
$c\bar{s}$ meson size, the strong couplings  to the hadronic
channels (or the virtual mesons loops) may result in smaller masses
than the conventional $c\bar{s}$ mesons in the constituent quark
models, and enrich the pure $c\bar{s}$ states with other components.
\end{abstract}


\section{Introduction}
 The two strange-charmed mesons $D_{s0} (2317)$ and $D_{s1} (2460)$ with the
 spin-parity $0^+$ and $1^+$
respectively can not be comfortably identified as the
quark-antiquark bound states in the spectrum of the constituent
quark models, they have triggered hot debate on their nature,
under-structures and whether it is necessary to introduce the exotic
states \cite{exp03,Swanson06}. The masses of the $D_{s0}(2317)$ and
$D_{s1}(2460)$ are significantly lower than the masses of the $0^+$
and $1^+$ states  respectively from the quark models and lattice
simulations \cite{QuarkLattice}. The difficulties to identify the
$D_{s0}(2317)$ and $D_{s1}(2460)$ states with the conventional
$c\overline{s}$ mesons are rather similar to those appearing in the
light scalar mesons below $1GeV$. The light scalar mesons are the
subject of an intense and continuous controversy in clarifying the
hadron spectroscopy\cite{Godfray}, the more elusive things are the
constituent structures of the $f_0(980)$ and $a_0(980)$ mesons with
almost the degenerate masses. The mesons $D_{s0} (2317)$ and $D_{s1}
(2460)$ lie just below the $D K$ and $D^\ast K$ threshold
respectively,  which are analogous to the situation that the scalar
mesons  $a_0(980)$ and $f_0(980)$ lie just below the $K\bar{K}$
threshold and  couple strongly to the nearby channels. The mechanism
responsible for the low-mass charmed mesons may be the same as the
light scalar nonet mesons, the $f_{0}(600)$, $f_{0}(980)$,
$a_{0}(980)$  and $K^{\ast}_{0}(800)$
\cite{ColangeloWang,WangWan06,ReviewScalar,WangScalar05}. There have
been a lot of explanations for their nature, for example, the
conventional $c\bar{s}$ states \cite{2quark,Colangelo2005},
two-meson molecular states \cite{2meson}, four-quark states
\cite{4quark}, etc. If we take the scalar mesons  $a_0(980)$ and
$f_0(980)$ as four quark states with the
 constituents  of scalar diquark-antidiquark  sub-structures, the
masses of the scalar nonet mesons below  $1GeV$ can be naturally
explained \cite{ReviewScalar,WangScalar05}.

 There are other possibilities besides the four-quark state explanations, for
example, the scalar mesons $a_0(980)$, $f_0(980)$, $D_{s0}(2317)$
and the axial-vector meson $D_{s1}(2460)$ may have bare $P-$wave
$q\overline{q}$ and $c\bar{s}$ kernels   with strong coupling to the
nearby thresholds respectively, the $S-$wave  virtual intermediate
hadronic states (or the virtual mesons loops) play a crucial role in
the composition of those bound states (or resonances due to the
masses below or above the thresholds). The hadronic dressing
mechanism (or unitarized quark models) takes the point of view that
the mesons $f_0(980)$, $a_0(980)$, $D_{s0}(2317)$ and $D_{s1}(2460)$
have small $q\bar{q}$ and $c\bar{s}$  kernels of the typical
$q\bar{q}$ and $c\bar{s}$  mesons size respectively. The strong
couplings to the virtual intermediate hadronic states (or the
virtual mesons loops) may result in smaller masses than the
conventional scalar $q\bar{q}$   and $c\bar{s}$  mesons in the
constituent quark models, enrich the pure $q\bar{q}$ and $c\bar{s}$
states with other components \cite{HDress,UQM}. Those mesons may
spend part (or most part) of their lifetime as virtual $ K \bar{K}
$, $DK$ and $D^*K$ states \cite{ColangeloWang,WangWan06,HDress,UQM}.
Despite what constituents they may have, we have the fact that  they
 lie just a little below the $K\bar{K}$, $DK$ and $D^*K$ thresholds respectively, the
strong interactions with the $K\bar{K}$, $DK$ and $D^*K$ thresholds
will significantly influence their dynamics, although the decays
$D_{s0}(2317)\rightarrow DK$ and $D_{s1}(2460)\rightarrow D^*K$ are
kinematically suppressed.  It is interesting to investigate the
possibility of the hadronic dressing mechanism.

In our previous work, we take the point of view that the scalar
mesons $f_0(980)$, $a_0(980)$ and $D_{s0}(2317)$ are the
conventional $q\bar{q}$ and $c\bar{s}$ state respectively, and
calculate the values of the strong coupling constants $g_{f_0KK}$,
$g_{a_0KK}$, and $g_{D_{s0} DK}$ within the framework of the
light-cone QCD sum rules approach \cite{ColangeloWang,WangWan06}.
The large values of the strong coupling constants support the
hadronic dressing mechanism. In this article, we take the
axial-vector meson $D_{s1}(2460)$ as the conventional $c\bar{s}$
state, and calculate the value of the strong coupling constant
$g_{D_{s1} D^*K}$ in the framework of the light-cone QCD sum rules
approach and study the possibility of the hadronic dressing
mechanism in the axial-vector channel. The light-cone QCD sum rules
approach carries out the operator product expansion near the
light-cone $x^2\approx 0$ instead of the short distance $x\approx 0$
while the non-perturbative matrix elements  are parameterized by the
light-cone distribution amplitudes
 which classified according to their twists  instead of
 the vacuum condensates \cite{LCSR,LCSRreview}. The non-perturbative
 parameters in the light-cone distribution amplitudes are calculated by   the conventional QCD  sum rules
 and the  values are universal \cite{SVZ79}.

The article is arranged as: in Section 2, we derive the strong
coupling constant  $g_{D_{s1} D^* K}$ within the framework of the
light-cone QCD sum rules approach; in Section 3, the numerical
result and discussion; and in Section 4, conclusion.

\section{Strong coupling constant  $g_{D_{s1} D^* K}$  with light-cone QCD sum rules}

In the following, we write down the definition  for the strong
coupling constant $g_{D_{s1} D^* K}$ ,
\begin{eqnarray}
\langle D_{s1}(p+q)  | D^*(p)K(q)\rangle=-ig_{D_{s1} D^*
K}\eta^*_\alpha \epsilon^\alpha=-iM_{D_{s1}}\hat{g}_{D_{s1} D^* K}
\eta^*_\alpha \epsilon^\alpha\, \, ,
\end{eqnarray}
where the $\epsilon_\alpha$ and $\eta_\alpha$ are the polarization
vectors of the mesons $D^*$ and $D_{s1}(2460)$ respectively. The
mass  of the $D_{s1}(2460)$ $M_{D_{s1}}$ can serve as an energy
scale, we factorize the $M_{D_{s1}}$ from the $g_{D_{s1} D^* K}$. We
study the strong coupling constant $g_{D_{s1} D^* K}$ with the
 two-point correlation function $\Pi_{\mu\nu}(p,q)$,
\begin{eqnarray}
\Pi_{\mu \nu}(p,q)&=&i \int d^4x \, e^{-i q \cdot x} \,
\langle 0 |T\left\{J^V_\mu(0) {J^{A}_{\nu}}^+(x)\right\}|K(p)\rangle \, , \\
J^V_\mu(x)&=&{\bar u}(x)\gamma_\mu  c(x)\, ,  \\
J^{A}_\mu(x)&=&{\bar s}(x)\gamma_\mu \gamma_5 c(x)\, ,
\end{eqnarray}
where the vector current $J^V_\mu(x)$ and the axial-vector current
$J^A_\mu(x)$ interpolate the vector meson $D^*$  and the
axial-vector meson $D_{s1}(2460)$ respectively, the external $K$
state has four momentum $p_\mu$ with $p^2=m_K^2$. The correlation
function $\Pi_{\mu\nu}(p,q)$ can be decomposed as
\begin{eqnarray}
\Pi_{\mu \nu}(p,q)&=&i \Pi g_{\mu\nu}+\Pi_1(p_\mu q_\nu+p_\nu q_\mu)
+\cdots
\end{eqnarray}
due to the Lorentz invariance.

According to the basic assumption of current-hadron duality in the
QCD sum rules approach \cite{SVZ79}, we can insert  a complete
series of intermediate states with the same quantum numbers as the
current operators $J^V_\mu(x)$ and $J^A_\mu(x)$  into the
correlation function $\Pi_{\mu\nu}(p,q)$ to obtain the hadronic
representation. After isolating the ground state contributions from
the pole terms of the mesons $D_{s1}(2460)$ and $D^*$, we get the
following result,
\begin{eqnarray}
\Pi_{\mu\nu}&=&\frac{\langle0| J^V_{\mu}(0)\mid
D^*(q+p)\rangle\langle D^*| D_{s1}K\rangle  \langle
D_{s1}(q)|{J^A_{\nu}}^+(0)| 0\rangle}
  {\left[M_{D^*}^2-(q+p)^2\right]\left[M_{D_{s1}}^2-q^2\right]}  + \cdots \nonumber \\
&=&-\frac{i g_{D_{s1}D^*K} f_{D^*} f_{D_{s1}}  M_{D^*}M_{D_{s1}}}
  {\left[M_{D^*}^2-(q+p)^2\right]\left[M_{D_{s1}}^2-q^2\right]}g_{\mu\nu} + \cdots  ,
\end{eqnarray}
where the following definitions have been used,
\begin{eqnarray}
\langle0 | J^V_{\mu}(0)|D^*\rangle&=&f_{D^*}M_{D^*}\epsilon_\mu\,, \nonumber\\
\langle0 |
J^A_{\mu}(0)|D_{s1}\rangle&=&f_{D_{s1}}M_{D_{s1}}\eta_\mu\,
,\nonumber
\end{eqnarray}
here the $f_{D^*}$ and $f_{D_{s1}}$ are the weak decay constants of
the $D^*$ and $D_{s1}(2460)$ respectively.
 The vector current $J^V_\mu(x)$ and axial-vector current
$J^A_\mu (x)$ have non-vanishing couplings to the scalar meson
$D_{0}$ and pseudoscalar meson $D_s$,  respectively,
\begin{eqnarray}
\langle0|J^V_\mu(0)|D_0(q)\rangle&=& f_{D_0}q_\mu \, ,\nonumber \\
\langle0|J^A_\mu(0)|D_s(q)\rangle&=& if_{D_s} q_\mu\, ,\nonumber
\end{eqnarray}
where the $f_{D_0}$ and $f_{D_s}$ are the weak decay constants. The
$\Pi_1$ with the tensor structure $p_\mu q_\nu+p_\nu q_\mu$ receives
contribution from the mesons $D_{0}$ and $D_s$ besides the $D^*$ and
$D_{s1}(2460)$, we choose the tensor structure $g_{\mu\nu}$ for
analysis  to avoid possible contaminations from the scalar and
pseudoscalar mesons.
 In Eq.(6), we have not shown the contributions from the high
resonances and continuum states explicitly as they are suppressed
due to the double Borel transformation.

In the following, we briefly outline the  operator product expansion
for the correlation function $\Pi_{\mu \nu}(p,q)$ in perturbative
QCD theory. The calculations are performed at the large space-like
momentum regions $(q+p)^2\ll 0$  and  $q^2\ll 0$, which correspond
to the small light-cone distance $x^2\approx 0$ required by the
validity of the operator product expansion approach. We write down
the propagator of a massive quark in the external gluon field in the
Fock-Schwinger gauge firstly \cite{Belyaev94},
\begin{eqnarray}
\langle 0 | T \{q_i(x_1)\, \bar{q}_j(x_2)\}| 0 \rangle &=&
 i \int\frac{d^4k}{(2\pi)^4}e^{-ik(x_1-x_2)}\nonumber\\
 && \left\{
\frac{\not\!k +m}{k^2-m^2} \delta_{ij} -\int\limits_0^1 dv\, g_s \,
G^{\mu\nu}_{ij}(vx_1+(1-v)x_2) \right. \nonumber \\
&&\left. \Big[ \frac12 \frac {\not\!k
+m}{(k^2-m^2)^2}\sigma_{\mu\nu} - \frac1{k^2-m^2}v(x_1-x_2)_\mu
\gamma_\nu \Big]\right\}\, ,
\end{eqnarray}
where the $G_{\mu \nu }$ is the gluonic field strength, and the
$g_s$ denotes the strong coupling constant. Substituting the above
$c$ quark propagator and the corresponding $K$ meson light-cone
distribution amplitudes into the correlation function
$\Pi_{\mu\nu}(p,q)$ in Eq.(2) and completing the integrals over the
variables  $x$ and $k$, finally we obtain the analytical result,
which is given explicitly in the appendix.

In calculation, the  two-particle and three-particle $K$ meson
light-cone distribution amplitudes have been used
\cite{LCSR,LCSRreview,Belyaev94,Ball98,Ball06}, the explicitly
expressions are given in the appendix. The parameters in the
light-cone distribution amplitudes are scale dependent and can be
estimated with the QCD sum rules approach
\cite{LCSR,LCSRreview,Belyaev94,Ball98,Ball06}. In this article, the
energy scale $\mu$ is chosen to be  $\mu=1GeV$.

Now we perform the double Borel transformation with respect to  the
variables $Q_1^2=-q^2$ and  $Q_2^2=-(p+q)^2$ for the correlation
function $\Pi$ in Eq.(6),   and obtain the analytical expression of
the invariant function in the hadronic representation,
\begin{eqnarray}
B_{M_2^2}B_{M_1^2}\Pi&=&-i\frac{
g_{D_{s1}D^*K}f_{D^*}f_{D_{s1}}M_{D^*}M_{D_{s1}}}{M_1^2M_2^2}
\exp\left[-\frac{M^2_{D_{s1}}}{M_1^2}
-\frac{M^2_{D^*}}{M_2^2}\right] +\cdots,
\end{eqnarray}
here we have not shown  the contributions from the high resonances
and continuum states  explicitly for simplicity. In order to match
the duality regions below the thresholds $s_0$ and $s_0'$ for the
interpolating currents $J^{V}_\mu(x)$   and  $J^A_\mu(x)$
respectively, we can express the correlation function $\Pi$ at the
level of quark-gluon degrees of freedom into the following form,
\begin{eqnarray}
\Pi&=& \int ds ds^\prime \frac{\rho(s,s^\prime)}{
[s-(q+p)^2][s'-q^2]} \, ,
\end{eqnarray}
then perform the double Borel transformation with respect to the
variables $Q_1^2$ and $Q_2^2$  directly. However, the analytical
expression of the spectral density $\rho(s,s')$ is hard to obtain,
we have to resort to  some approximations.  As the contributions
 from the higher twist terms  are  suppressed by more powers of
 $\frac{1}{m_c^2-(q+up)^2}$, the net contributions of the three-particle
(quark-antiquark-gluon) twist-3 and twist-4  terms are of minor
importance, about $20\%$, the continuum subtractions will not affect
the results remarkably. The dominating contribution comes from the
two-particle twist-3 term involving the $\varphi_p(u)$. We preform
the same trick as Refs.\cite{Belyaev94,Kim} and expand the amplitude
$\varphi_p(u)$ in terms of polynomials of $1-u$,
\begin{eqnarray}
\varphi_p(u)=\sum_{k=0}^N b_k(1-u)^k=\sum_{k=0}^N b_k
\left(\frac{s-m_c^2}{s-q^2}\right)^k,
\end{eqnarray}
then introduce the variable $s'$ and the spectral density is
obtained. In the decay $B \to \chi_{c0}K$, the factorizable
contribution is zero and the non-factorizable contributions from the
soft hadronic matrix elements are too small to accommodate the
experimental data \cite{WangLH}, the contributions of the
three-particle (quark-antiquark-gluon)  distribution amplitudes of
the mesons are always of minor importance comparing with the
two-particle (quark-antiquark) distribution amplitudes in the
light-cone QCD sum rules. In our previous work, we  study the four
form-factors $f_1(Q^2)$, $f_2(Q^2)$, $g_1(Q^2)$ and $g_2(Q^2)$ of
the $\Sigma \to n$ in the framework of the light-cone QCD sum rules
approach up to twist-6 three-quark light-cone distribution
amplitudes and obtain satisfactory results \cite{Wang06}. In the
light-cone QCD sum rules,
 we can neglect the contributions from the valence gluons and make relatively rough estimations.

After straightforward calculations, we obtain the final expression
of the double Borel transformed correlation function $\Pi
(M_1^2,M_2^2)$ at the level of quark-gluon degrees of freedom. The
masses of  the charmed mesons are $M_{D_{s1}}=2.46GeV$ and
$M_{D^*}=2.01GeV$, $\frac{M_{D^*}}{M_{D^*}+M_{D_{s1}}}\approx0.45$,
 there exists an overlapping working window for the two Borel
parameters $M_1^2$ and $M_2^2$, it's convenient to take the value
$M_1^2=M_2^2$. We introduce the threshold parameter $s_0$ and make
the simple replacement,
\begin{eqnarray}
e^{-\frac{m_c^2+u_0(1-u_0)m_K^2}{M^2}} \rightarrow
e^{-\frac{m_c^2+u_0(1-u_0)m_K^2}{M^2} }-e^{-\frac{s_0}{M^2}}
\nonumber
\end{eqnarray}
 to subtract the contributions from the high resonances  and
  continuum states \cite{Belyaev94}, finally we obtain the sum rule for the strong coupling constant
$g_{D_{s1}D^*K}$,
\begin{eqnarray}
g_{D_{s1}D^*K}&=& \frac{1}{f_{D^*}f_{D_{s1}}M_{D^*}M_{D_{s1}}}
\exp\left( \frac{M^2_{D_{s1}}}{M_1^2} +\frac{M^2_{D^*}}{M_2^2}
\right) \left\{\left[\exp\left(- \frac{BB}{M^2}\right)-\exp\left(-
\frac{s_0}{M^2}\right)\right] \right.\nonumber\\
 &&f_K\left[\frac{ m_c m_K^2M^2}{m_u+m_s}
\varphi_p(u_0)
+\frac{m_K^2(M^2+m_c^2)}{8}  \frac{d}{du_0}A(u_0) -\frac{M^4}{2} \frac{d}{du_0}\phi_K(u_0)\right] \nonumber\\
&&-\exp\left(-\frac{BB}{M^2}\right) \left[f_K m_c^2 m_K^2
\int_0^{u_0} dt B(t) \right.\nonumber\\
&&+ m_K^2 \int_0^{u_0} d\alpha_s
\int_{u_0-\alpha_s}^{1-\alpha_s} d\alpha_g \frac{(u_0f_Km_K^2\Phi+f_{3K}m_c \phi_{3K})(1-\alpha_s-\alpha_g,\alpha_s,\alpha_g)}{\alpha_g} \nonumber \\
&&+f_Km_K^2M^2\frac{d}{du_0}\int_0^{u_0} d\alpha_s
\int_{u_0-\alpha_s}^{1-\alpha_s} d\alpha_g
\frac{(A_\parallel-V_\parallel)(1-\alpha_s-\alpha_g,\alpha_s,\alpha_g)}{2\alpha_g}
 \nonumber \\
&&-f_Km_K^2M^2\frac{d}{du_0}\int_0^{u_0} d\alpha_s
\int_{u_0-\alpha_s}^{1-\alpha_s} d\alpha_g
A_\parallel(1-\alpha_s-\alpha_g,\alpha_s,\alpha_g)\frac{\alpha_s+\alpha_g-u_0}{\alpha_g^2}
 \nonumber \\
 &&+f_Km_K^4  \left(\int_0^{1-u_0} d\alpha_g \int^{u_0}_{u_0-\alpha_g}
d\alpha_s \int_0^{\alpha_s} d\alpha +\int^1_{1-u_0} d\alpha_g
\int^{1-\alpha_g}_{u_0-\alpha_g}
d\alpha_s \int_0^{\alpha_s} d\alpha\right)  \nonumber \\
&& \left[\frac{1}{\alpha_g}\left(3-\frac{2m_c^2}{M^2}\right)\Phi+\frac{4m_c^2}{M^2}\frac{\alpha_s+\alpha_g-u_0}{\alpha_g^2}(A_\perp+A_\parallel)\right](1-\alpha-\alpha_g,\alpha,\alpha_g)\nonumber \\
 &&-f_Km_K^4  u_0\frac{d}{du_0}\left(\int_0^{1-u_0} d\alpha_g \int^{u_0}_{u_0-\alpha_g}
d\alpha_s \int_0^{\alpha_s} d\alpha +\int^1_{1-u_0} d\alpha_g
\int^{1-\alpha_g}_{u_0-\alpha_g}
d\alpha_s \int_0^{\alpha_s} d\alpha\right)  \nonumber \\
&& \frac{\Phi(1-\alpha-\alpha_g,\alpha,\alpha_g)}{\alpha_g} \nonumber \\
&&-f_Km_K^4   \int_{1-u_0}^1 d\alpha_g \int_0^{\alpha_g} d\beta
\int_0^{1-\beta} d\alpha \left[\Phi(1-\alpha-\beta,\alpha,\beta)
\frac{1-u_0}{\alpha_g^2}
 \left(4-\frac{2m_c^2}{M^2}\right) \right.\nonumber \\
&&\left. +\frac{4m_c^2}{M^2}\frac{(1-u_0)^2}{\alpha_g^3}(A_\parallel+A_\perp)(1-\alpha-\beta,\alpha,\beta)\right]\nonumber \\
&&\left.\left.+f_Km_K^4  \frac{d}{du_0} \int_{1-u_0}^1 d\alpha_g
\int_0^{\alpha_g} d\beta \int_0^{1-\beta} d\alpha
\Phi(1-\alpha-\beta,\alpha,\beta) \frac{u_0(1-u_0)}{\alpha_g^2}
\right]\right\} ,
\end{eqnarray}
where
\begin{eqnarray}
BB&=&m_c^2+u_0(1-u_0)m_K^2 \, ,\nonumber \\
u_0&=&\frac{M_1^2}{M_1^2+M_2^2}\, , \nonumber \\
M^2&=&\frac{M_1^2M_2^2}{M_1^2+M_2^2} \, ,
\end{eqnarray}
here we write down only the analytical result without the technical
details
  \footnote{ In this footnote, we present some technical details
  necessary  in performing the Borel transformation which are not familiar to the
   novices,
\begin{eqnarray}
&&\int_0^1 dv \int_0^1 d\alpha_g \int_0^{1-\alpha_g} d\alpha_s
f(v,\alpha_s,\alpha_g) \frac{d}{du}
\exp\left[-\frac{m_c^2+u(1-u)m_K^2}{M^2}
\right]\delta(u-u_0)|_{u=\alpha_s+(1-v)\alpha_g} \nonumber \\
&=&\int_0^1 du\int_0^1 dv \int_0^1 d\alpha_g \int_0^{1-\alpha_g}
d\alpha_s f(v,\alpha_s,\alpha_g)
\delta\left[u-\alpha_s-(1-v)\alpha_g\right]\nonumber\\
&&\frac{d}{du} \exp\left[-\frac{m_c^2+u(1-u)m_K^2}{M^2}
\right]\delta(u-u_0) \nonumber \\
&=&\int_0^1 du \int_0^u d\alpha_s \int_{u-\alpha_s}^{1-\alpha_s}
d\alpha_g \frac{f(v,\alpha_s,\alpha_g)}{\alpha_g} \frac{d}{du}
\exp\left[-\frac{m_c^2+u(1-u)m_K^2}{M^2}
\right]\delta(u-u_0) \nonumber \\
&=&-\int_0^1 du \exp\left[-\frac{m_c^2+u(1-u)m_K^2}{M^2}
\right]\delta(u-u_0)\frac{d}{du}\int_0^u d\alpha_s
\int_{u-\alpha_s}^{1-\alpha_s} d\alpha_g
\frac{f(v,\alpha_s,\alpha_g)}{\alpha_g}
 \nonumber \\
 &=&- \exp\left[-\frac{m_c^2+u_0(1-u_0)m_K^2}{M^2}
\right]\frac{d}{du_0}\int_0^{u_0} d\alpha_s
\int_{u_0-\alpha_s}^{1-\alpha_s} d\alpha_g
\frac{f(v,\alpha_s,\alpha_g)}{\alpha_g} \, ,
 \nonumber
\end{eqnarray}
where the $f(v,\alpha_s,\alpha_g)$ stands for the three-particle
light-cone distribution amplitudes.
  }. The term proportional to the $M^4\frac{d}{du_0}\phi_K(u_0)$ in Eq.(11)  depends heavily on the asymmetry
  coefficient  $a_1(\mu)$ of the twist-2 light-cone distribution amplitude
  $\phi_K(u)$ in the  limit  $u_0=\frac{1}{2}$, if we take the value $a_1(\mu)=0.06\pm 0.03$ \cite{Ball98,Ball06},
  no stable sum rules can be obtained, the value of the $g_{D_{s1}D^*K}$ changes
  significantly with the variation of the Borel
  parameter $M^2$.  In this article, we take the assumption that the $u$ and $s$ quarks
  have  symmetric momentum distributions and  neglect the coefficient $a_1(\mu)$.
  The existence of such a term is not a bad thing.
  The $D_{s1}(2460)$ lies below the $D^*K$ threshold, it is
  impossible  to measure the strong coupling constant
  $g_{D_{s1}D^*K}$ directly, the corresponding beauty doublet $(0^+,1^+)$
  $B_{s0}$ and $B_{s1}$ may lie above the $BK$ and $B^*K$
  thresholds respectively. Once the experimental data of the
  $B_{s0}$, $B_{s1}$ and the related strong coupling constants are
  available, we can compare  the values of the $g_{B_{s1}B^*K}$ from
  the QCD sum rules with the ones  from the experiment, and verify
  whether or not the coefficient $a_1(\mu)$ can be safely neglected.
  That will put severe constraint on the value of the $a_1(\mu)$.
  With the  simple replacement
  \begin{eqnarray}
  m_c &\rightarrow& m_b \, ,\nonumber \\
  M_{D_{s1}} &\rightarrow& M_{B_{s1}} \, ,\nonumber \\
   M_{D^*} &\rightarrow& M_{B^*} \, ,\nonumber \\
    f_{D_{s1}} &\rightarrow& f_{B_{s1}} \, ,\nonumber \\
   f_{D^*} &\rightarrow& f_{B^*}
  \end{eqnarray}
in Eq.(11), we can obtain the QCD sum rule for the strong coupling
constant $g_{B_{s1}B^*K}$.

\section{Numerical result and discussion}
The parameters are taken as $m_s=(140\pm 10 )MeV$,
$m_q=(5.6\pm1.6)MeV$, $m_c=(1.25\pm 0.10)GeV$,
$\lambda_3=1.6\pm0.4$, $f_{3K}=(0.45\pm0.15)\times 10^{-2}GeV^2$,
$\omega_3=-1.2\pm0.7$, $\eta_4=0.6 \pm0.2 $, $\omega_4=0.2\pm0.1$,
$a_2=0.25\pm 0.15$ \cite{LCSR,LCSRreview,Belyaev94,Ball98,Ball06},
$f_K=0.160GeV$, $m_K=498MeV$, $M_{D_{s1}}(2460)=2.46GeV$,
$M_{D^*}=2.01GeV$, $f_{D^*}=(0.24\pm0.02)GeV$ \cite{Belyaev94}, and
$f_{D_{s1}}=(0.225\pm0.020)GeV$ \cite{Colangelo2005}. The duality
threshold $s_0$ in Eq.(11) is taken as $s_0=(6.8-7.2)GeV^2$
 to avoid possible  contaminations from the high resonances and
continuum states, it is reasonable for the narrow $D_{s1}(2460)$
$\sqrt{s_0}=(2.6-2.7)GeV>M_{D_{s1}}$, furthermore, in this region,
the numerical result is not sensitive to the threshold parameter
$s_0$. The Borel parameters are chosen as $ M_1^2=M_2^2=
(6-14)GeV^2$ and $M^2=(3-7)GeV^2$, in those regions, the value of
the strong coupling constant $g_{D_{s1} D^*K}$ is rather stable from
the sum rule in Eq.(11) with the simple subtraction, which is shown
in  Figs.(1-2).

The uncertainties of the four parameters  $m_u$,  $m_c$, $\lambda_3$
and $\omega_3$  can only result in small uncertainties for the
numerical values. The main uncertainties come from the seven
parameters
 $f_{3K}$, $m_s$, $f_{D^*}$,
$f_{D_{s1}}$,  $a_2$, $\eta_4$ and $\omega_4$,  the variations of
those parameters (except for the $a_2$) can lead to  large changes
for the numerical values, about $(5-10)\%$, which are shown in the
Fig.1. The uncertainties of the three hadronic parameters $f_{3K}$,
$f_{D^*}$, $f_{D_{s1}}$ and the two light-cone distribution
parameters $\eta_4$, $\omega_4$ can be pined down with the improved
QCD sum rules or more experimental data, however, it is a difficult
work.
\begin{figure}
\centering
  \includegraphics[totalheight=6cm,width=7cm]{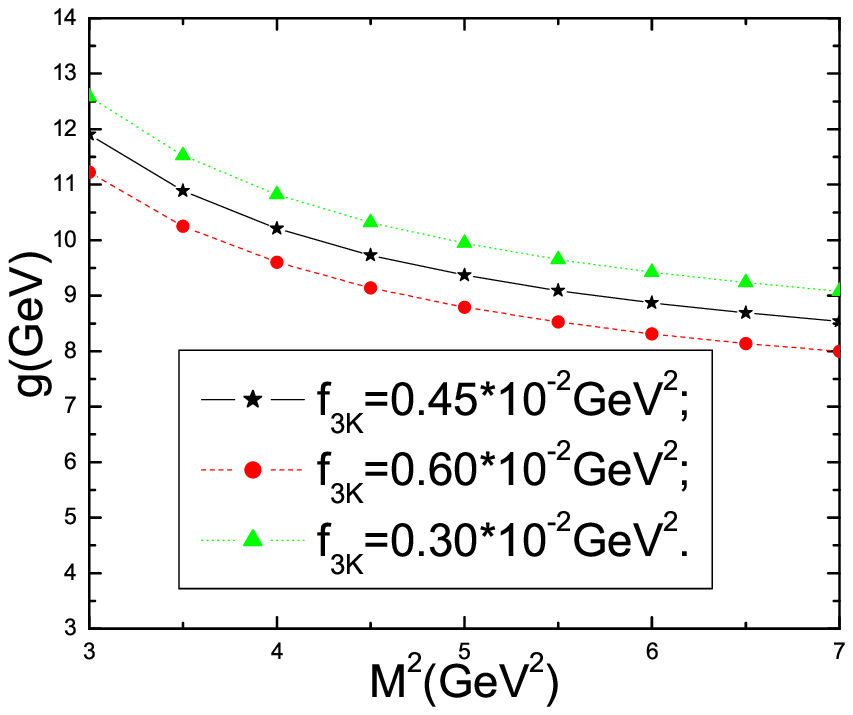}
  \includegraphics[totalheight=6cm,width=7cm]{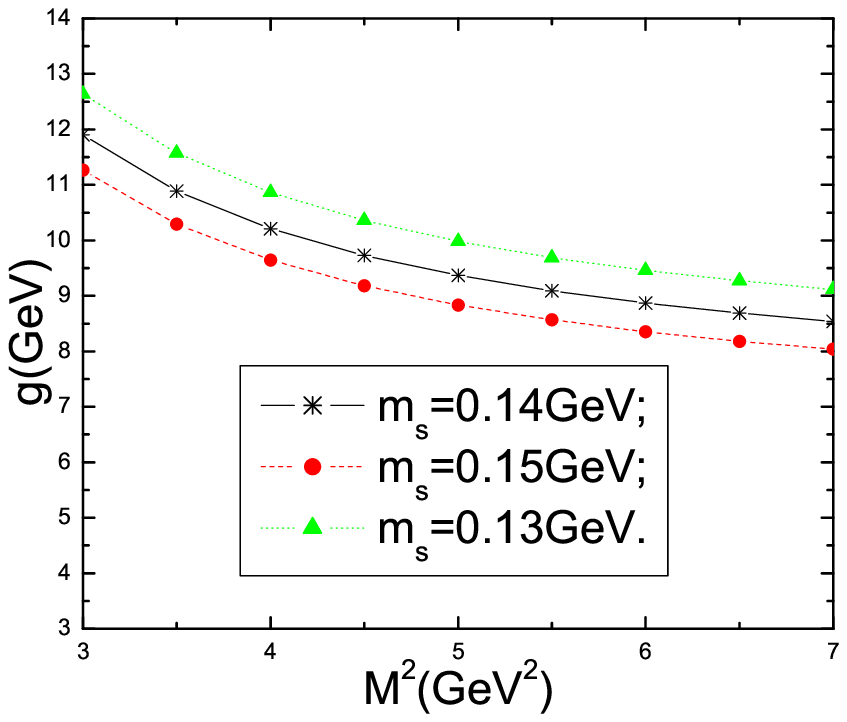}
  \includegraphics[totalheight=6cm,width=7cm]{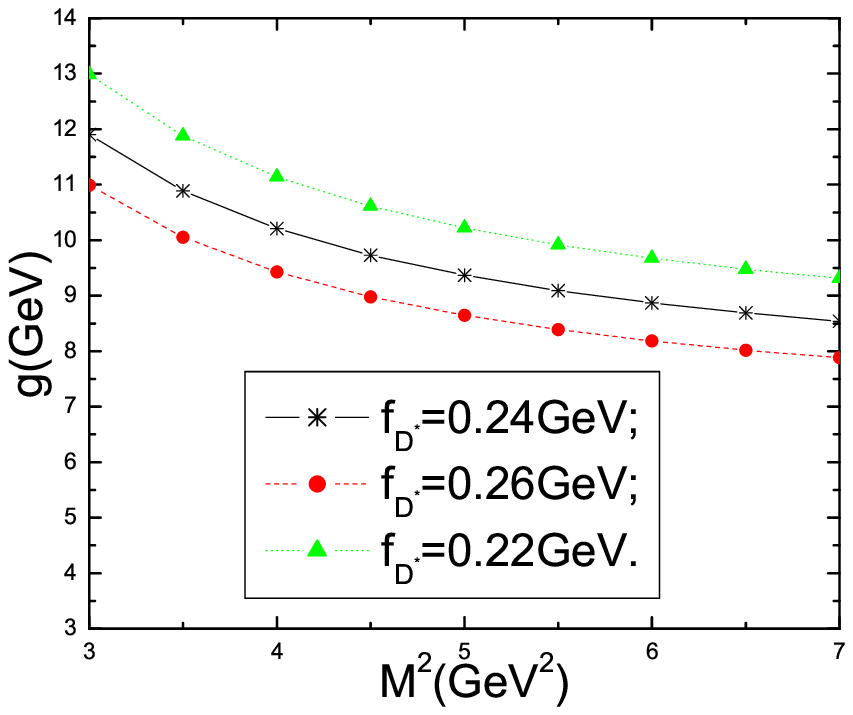}
  \includegraphics[totalheight=6cm,width=7cm]{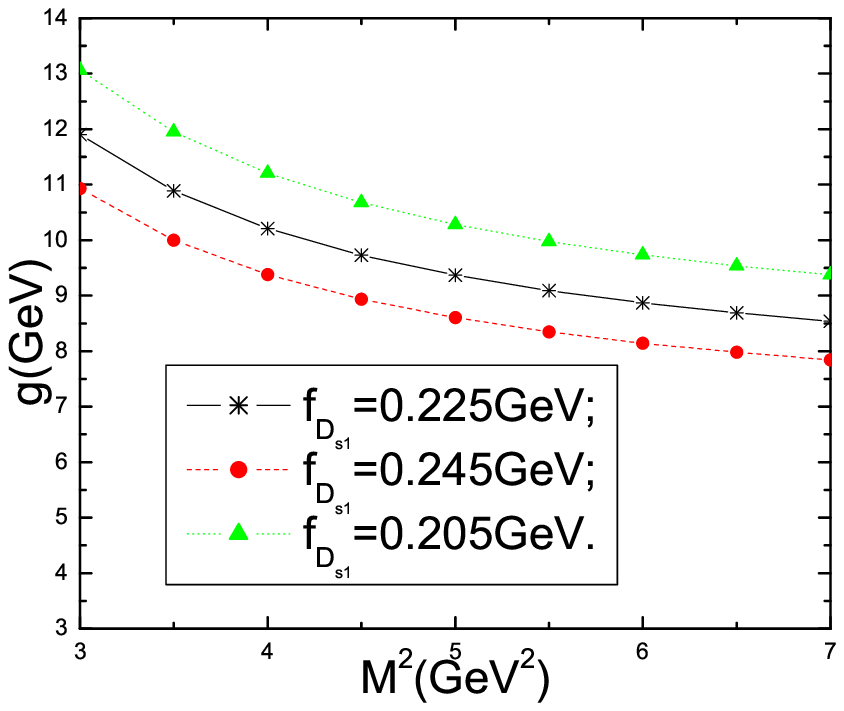}
  \includegraphics[totalheight=6cm,width=7cm]{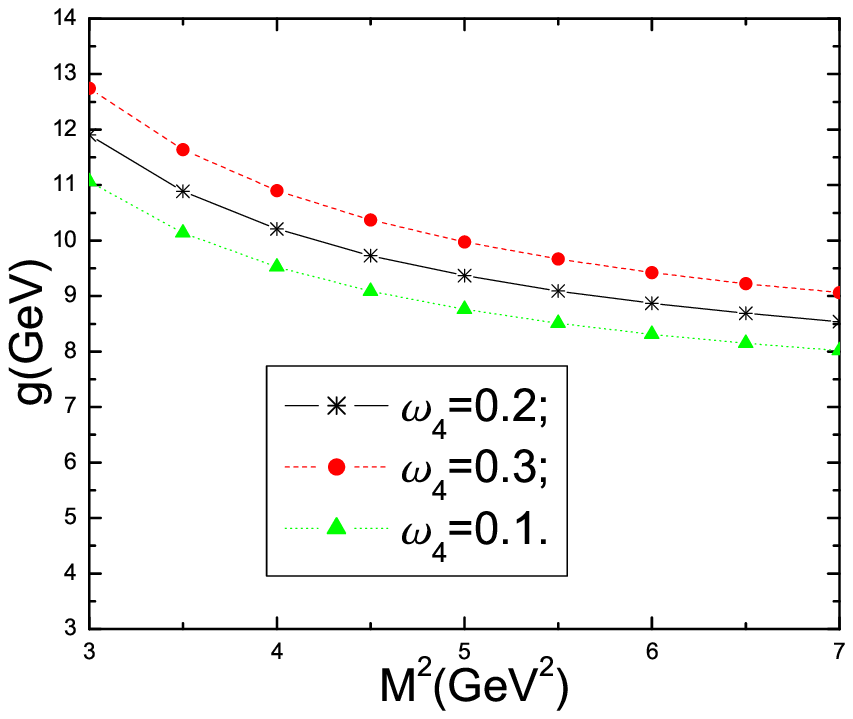}
  \includegraphics[totalheight=6cm,width=7cm]{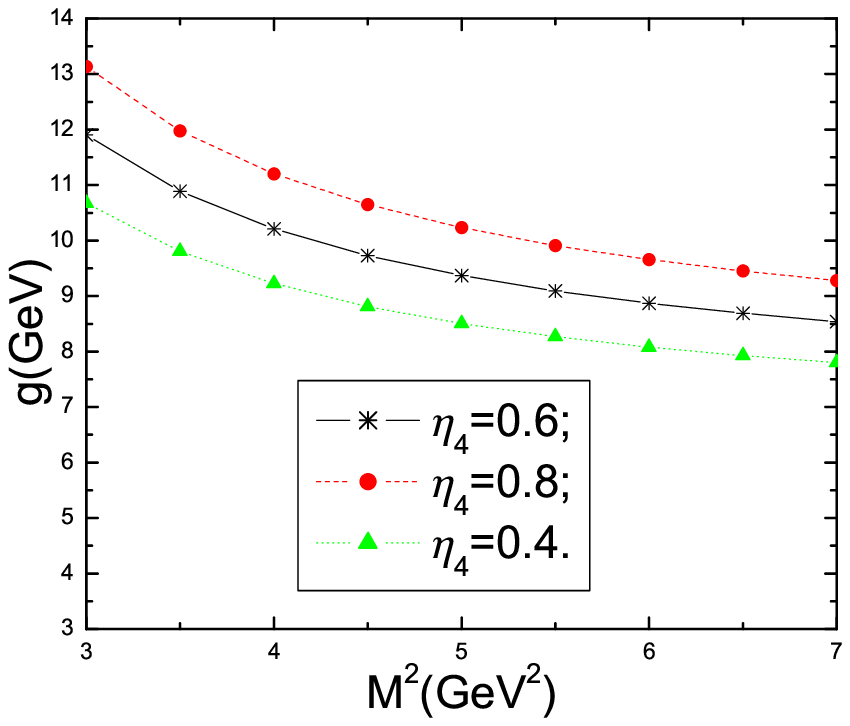}
     \caption{The   $g_{D_{s1}D^*K}$ with the parameters $f_{3K}$, $m_s$, $f_{D^*}$,
$f_{D_{s1}}$,  $\omega_4$    and $\eta_4$ respectively. }
\end{figure}
\begin{figure}
\centering
  \includegraphics[totalheight=9cm,width=13cm]{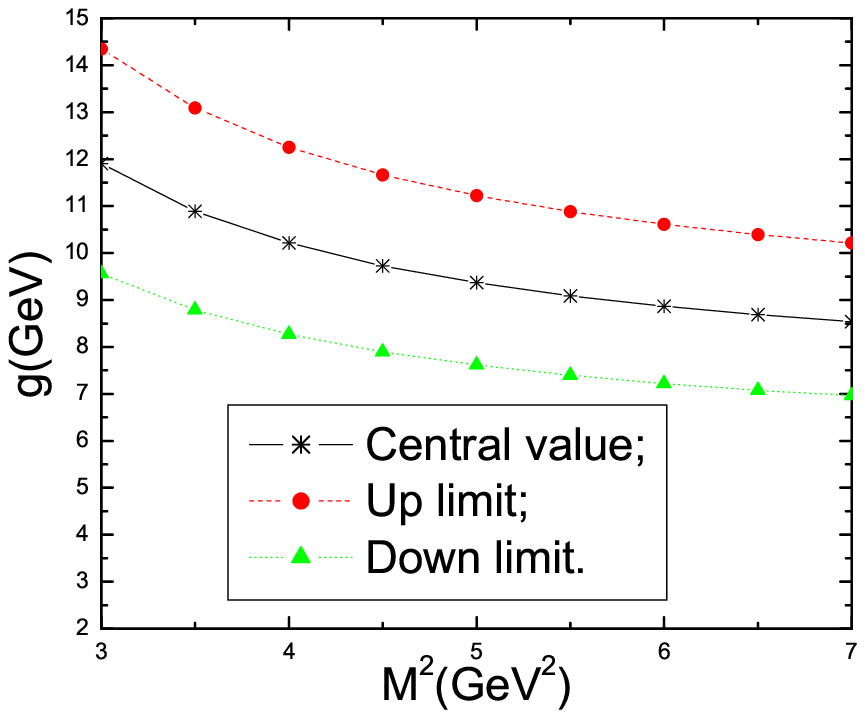}
     \caption{The   $g_{D_{s1}D^*K}$  with the Borel parameter $M^2$. }
\end{figure}

Taking into account all the uncertainties from the thirteen
parameters $m_s$, $m_q$, $m_c$, $\lambda_3$, $f_{3K}$, $\omega_3$,
$\eta_4$, $\omega_4$, $a_2$, $f_{D^*}$, $f_{D_{s1}}$, $s_0$ and
$M^2$, finally we obtain the numerical result of the strong coupling
constant,
\begin{eqnarray}
  g_{D_{s1}D^*K} &=&(10.5\pm3.5) GeV \, , \\
 \hat{g}_{D_{s1}D^*K} &=&4.3 \pm 1.4  \, ,
\end{eqnarray}
the uncertainty is large, about $30\%$. The large values of the
strong coupling constants $g_{D_{s0}DK}$ ($g_{D_{s0}DK} =(9.6 \pm
2.4) GeV$ \cite{WangWan06}) and $g_{D_{s1}D^*K}$ obviously support
  the hadronic
dressing mechanism \footnote{Here we will take a short discussion
about the hadronic dressing mechanism \cite{HDress,UQM}, one can
consult the original literatures  for the details. In the
conventional constituent quark models, the mesons are taken as
quark-antiquark bound states. The spectrum can be obtained by
solving the corresponding Schrodinger's or Dirac's equations with
the phenomenological potential which trying to incorporate the
observed properties of the strong interactions, such as the
asymptotic freedom and confinement. The solutions can be referred as
confinement bound states or bare quark-antiquark states (or
kernels). If we switch on the hadronic interactions between the
confinement bound states and the free ordinary  two-meson states,
the situation becomes more complex. With the increasing hadronic
coupling constants, the contributions from the hadronic loops of the
intermediate mesons become larger and  the bare quark-antiquark
states can be distorted greatly. There may be double poles or
several poles   in the scattering amplitudes with the same quantum
number as the bare quark-antiquark kernels; some ones stem  from the
bare quark-antiquark kernels while the others originate from the
continuum states. The strong coupling may enrich the bare
quark-antiquark states with other components, for example, the
virtual mesons pairs, and  spend part (or most part) of their
lifetime as virtual mesons pairs.}, the scalar meson  $D_{s0}(2317)$
and axial-vector meson $D_{s1}(2460)$ (just like the scalar mesons
$f_0(980)$ and $a_0(980)$, see Ref.\cite{ColangeloWang}) can be
taken as having small scalar and axial-vector  $c\bar{s}$  kernels
of typical meson size with large virtual $S$-wave $DK$ and $D^*K$
cloud respectively. In Ref.\cite{Guo06}, the authors analyze the
unitarized two-meson scattering amplitudes from the heavy-light
chiral Lagrangian,  and observe that the scalar meson $D_{s0}(2317)$
and  axial-vector meson $D_{s1}(2460)$ appear as the bound state
poles with the strong coupling constants $g_{D_{s0}DK}=10.203GeV$
and $g_{D_{s1}D^*K}=10.762GeV$. Our numerical results $g_{D_{s0}DK}
=(9.6\pm 2.4) GeV$ and $g_{D_{s1}D^*K} =(10.5 \pm 3.5) GeV$ are
certainly reasonable and can make robust predictions. However, we
take the point of view that the meson $D_{s0}(2317)$
($D_{s1}(2460)$) be bound state in the sense that it appears below
the $DK$ ($D^*K$) threshold, its constituents may be the bare
$c\bar{s}$ state, the virtual  $DK$ ($D^*K$) pair and their mixing,
rather than the $DK$ ($D^*K$) bound state. In Ref.\cite{Dai06}, the
authors take the point of view that the $D_{s0}(2317)$ is the scalar
$c\bar{s}$ meson and calculate the mass $M_{D_{s0}}$ with the QCD
sum rules approach by taking into account the contribution of the
$DK$ continuum, the effects of  the $DK$ continuum can pull the mass
down remarkably, and the value of the $M_{D_{s0}}$ is in good
agreement with experimental data. Our numerical values of the strong
coupling constants $g_{D_{s0}DK}$ and $g_{D_{s1}D^*K}$ are
approximately equal, the spin symmetry of the heavy quarks works
rather  well, the contribution of the $D^*K$ continuum may pull the
mass $M_{D_{s1}}$ down remarkably. One can analyze the value of the
$M_{D_{s1}}$ in the framework of the QCD sum rules with the
axial-vector current $J^A_\mu(x)$ by including the contribution of
the $D^*K$ continuum.

\section{Conclusion}

In this article, we take the point of view that the  charmed mesons
$D_{s0}(2317)$ and $D_{s1}(2460)$ are the conventional $c\bar{s}$
mesons and calculate the strong coupling constant $g_{D_{s1} D^* K}$
within the framework of the light-cone QCD sum rules approach. The
numerical values of the strong coupling constants $g_{D_{s1} D^* K}$
and $g_{D_{s0} D K}$ are compatible with the existing estimations,
the large values support the hadronic dressing mechanism. The
uncertainty of the value of the $g_{D_{s1} D^* K}$ is large, about
$30\%$, it comes from the uncertainties of the thirteen parameters
$m_s$, $m_q$, $m_c$, $\lambda_3$, $f_{3K}$, $\omega_3$, $\eta_4$,
$\omega_4$, $a_2$, $f_{D^*}$, $f_{D_{s1}}$, $s_0$ and $M^2$, while
the main uncertainty comes from the seven parameters
 $f_{3K}$, $m_s$, $f_{D^*}$,
$f_{D_{s1}}$,  $a_2$, $\eta_4$ and $\omega_4$,  the variations of
those parameters (except for the $a_2$) can lead to  large changes
for the numerical values, about $(5-10)\%$, refining those
parameters is of great importance, improved QCD sum rules and more
experimental data may pin down the uncertainties.

 Just
like the scalar mesons $f_0(980)$ and $a_0(980)$, the scalar meson
$D_{s0}(2317)$ and the axial-vector meson $D_{s1}(2460)$ may have
small $c\bar{s}$ kernels  of typical $c\bar{s}$ meson size. The
strong couplings  to virtual intermediate hadronic states (or the
virtual mesons loops) can result in  smaller masses  than the
conventional
 $0^+$ and $1^+$ mesons in the constituent quark models, enrich the
pure $c\bar{s}$ states with other components. The $D_{s0}(2317)$ and
$D_{s1}(2460)$ may spend part (or most part) of their lifetimes as
virtual $ D K $ and $ D^* K $ states, respectively.

 \section*{Appendix}
 The analytical expression of the $\Pi$ at the level of the
 quark-gluon degrees of freedom,
\begin{eqnarray}
\Pi &=&-\frac{f_K m_c m_K^2}{m_u+m_s} \int_0^1 du  \frac{\varphi_p
(u)}{AA}+f_K m_c^2 m_K^2 \int_0^1 du \int_0^u dt \frac{B(t)}{AA^2} \nonumber\\
&&+\frac{f_K}{2} \int_0^1 du \left\{ \phi_K(u) \frac{d}{du}\log AA+
\frac{A(u)m_K^2}{4} \frac{d}{du} \left[
\frac{1}{AA} +\frac{m_c^2}{AA^2}\right]\right\} \nonumber \\
&&+m_K^2 \int_0^1 dv \int_0^1 d\alpha_g \int_0^{1-\alpha_g}
d\alpha_s  \nonumber\\
&&\frac{\left[f_Km_K^2u\Phi+f_{3K}m_c\phi_{3K}\right](1-\alpha_s-\alpha_g,\alpha_s,\alpha_g)}{AA^2} \mid_{u=\alpha_s+(1-v)\alpha_g} \nonumber \\
&&-\frac{f_Km_K^2}{2} \int_0^1 dv \int_0^1 d\alpha_g
\int_0^{1-\alpha_g}
d\alpha_s  \left[(1-2v)A_\parallel -V_\parallel\right](1-\alpha_s-\alpha_g,\alpha_s,\alpha_g) \nonumber\\
&& \frac{d}{du} \frac{1}{AA}\mid_{u=\alpha_s+(1-v)\alpha_g} \nonumber \\
&&+f_Km_K^4 \int_0^1 dv \int_0^1 d\alpha_g \int_0^{1-\alpha_g}
d\alpha_s \int_0^{\alpha_s} d\alpha \Phi(1-\alpha-\alpha_g,\alpha,\alpha_g)  \nonumber \\
&&\left\{ \frac{4}{AA^2}-\frac{4m_c^2}{AA^3}+u\frac{d}{du}
\frac{1}{AA^2} \right\}_{u=\alpha_s+(1-v)\alpha_g} \nonumber \\
&&-f_Km_K^4 \int_0^1 dvv \int_0^1 d\alpha_g \int_0^{\alpha_g}
d\beta \int_0^{1-\beta} d\alpha \Phi(1-\alpha-\beta,\alpha,\beta)  \nonumber \\
&&\left\{ \frac{4}{AA^2}-\frac{4m_c^2}{AA^3}+u\frac{d}{du}
\frac{1}{AA^2} \right\}_{u=1-v\alpha_g} \nonumber \\
&&+8f_Km_c^2m_K^4 \int_0^1 dv v\int_0^1 d\alpha_g
\int_0^{1-\alpha_g}
d\alpha_s \int_0^{\alpha_s} d\alpha(A_\parallel+A_\perp)(1-\alpha-\alpha_g,\alpha,\alpha_g)  \nonumber \\
&& \frac{1}{AA^3}\mid_{u=\alpha_s+(1-v)\alpha_g} \nonumber \\
&&-8f_K m_c^2m_K^4 \int_0^1 dvv^2 \int_0^1 d\alpha_g
\int_0^{\alpha_g}
d\beta \int_0^{1-\beta} d\alpha (A_\parallel+A_\perp)(1-\alpha-\beta,\alpha,\beta)  \nonumber \\
&& \frac{1}{AA^3} \mid_{u=1-v\alpha_g}  \,\, ,
\end{eqnarray}
where
\begin{eqnarray}
AA&=&m_c^2-(q+up)^2 \, ,\nonumber \\
\Phi&=&A_\parallel+A_\perp-V_\parallel-V_\perp\, .
\end{eqnarray}

 The light-cone distribution amplitudes of the $K$ meson are defined
 by
\begin{eqnarray}
\langle0| {\bar u} (0) \gamma_\mu \gamma_5 s(x) |K(p)\rangle& =& i
f_K p_\mu \int_0^1 du  e^{-i u p\cdot x}
\left\{\phi_K(u)+\frac{m_K^2x^2}{16}
A(u)\right\}\nonumber\\
&&+f_K m_K^2\frac{ix_\mu}{2p\cdot x}
\int_0^1 du  e^{-i u p \cdot x} B(u) \, , \nonumber\\
\langle0| {\bar u} (0) i \gamma_5 s(x) |K(p)\rangle &=& \frac{f_K
m_K^2}{
m_s+m_u}\int_0^1 du  e^{-i u p \cdot x} \varphi_p(u)  \, ,  \nonumber\\
\langle0| {\bar u} (0) \sigma_{\mu \nu} \gamma_5 s(x) |K(p)\rangle
&=&i(p_\mu x_\nu-p_\nu x_\mu)  \frac{f_K m_K^2}{6 (m_s+m_u)}
\int_0^1 du
e^{-i u p \cdot x} \varphi_\sigma(u) \, ,  \nonumber\\
\langle0| {\bar u} (0) \sigma_{\alpha \beta} \gamma_5 g_s G_{\mu
\nu}(v x)s(x) |K(p)\rangle&=& f_{3 K}\left\{(p_\mu p_\alpha
g^\bot_{\nu
\beta}-p_\nu p_\alpha g^\bot_{\mu \beta}) -(p_\mu p_\beta g^\bot_{\nu \alpha}\right.\nonumber\\
&&\left.-p_\nu p_\beta g^\bot_{\mu \alpha})\right\} \int {\cal
D}\alpha_i \phi_{3 K} (\alpha_i)
e^{-ip \cdot x(\alpha_s+v \alpha_g)} \, ,\nonumber\\
\langle0| {\bar u} (0) \gamma_{\mu} \gamma_5 g_s G_{\alpha
\beta}(vx)s(x) |K(p)\rangle&=&  p_\mu  \frac{p_\alpha
x_\beta-p_\beta x_\alpha}{p
\cdot x}f_Km_K^2\nonumber\\
&&\int{\cal D}\alpha_i A_{\parallel}(\alpha_i) e^{-ip\cdot
x(\alpha_s +v \alpha_g)}\nonumber \\
&&+ f_Km_K^2 (p_\beta g_{\alpha\mu}-p_\alpha
g_{\beta\mu})\nonumber\\
&&\int{\cal D}\alpha_i A_{\perp}(\alpha_i)
e^{-ip\cdot x(\alpha_s +v \alpha_g)} \, ,  \nonumber\\
\langle0| {\bar u} (0) \gamma_{\mu}  g_s \tilde G_{\alpha
\beta}(vx)s(x) |K(p)\rangle&=& p_\mu  \frac{p_\alpha x_\beta-p_\beta
x_\alpha}{p \cdot
x}f_Km_K^2\nonumber\\
&&\int{\cal D}\alpha_i V_{\parallel}(\alpha_i) e^{-ip\cdot
x(\alpha_s +v \alpha_g)}\nonumber \\
&&+ f_Km_K^2 (p_\beta g_{\alpha\mu}-p_\alpha
g_{\beta\mu})\nonumber\\
&&\int{\cal D}\alpha_i V_{\perp}(\alpha_i) e^{-ip\cdot x(\alpha_s +v
\alpha_g)} \, ,
\end{eqnarray}
where the operator $\tilde G_{\alpha \beta}$  is the dual of the
$G_{\alpha \beta}$, $\tilde G_{\alpha \beta}= {1\over 2}
\epsilon_{\alpha \beta  \mu\nu} G^{\mu\nu} $ and ${\cal{D}}\alpha_i$
is defined as ${\cal{D}} \alpha_i =d \alpha_1 d \alpha_2 d \alpha_3
\delta(1-\alpha_1 -\alpha_2 -\alpha_3)$. The  light-cone
distribution amplitudes are parameterized as
\begin{eqnarray}
\phi_K(u,\mu)&=&6u(1-u)
\left\{1+a_1C^{\frac{3}{2}}_1(2u-1)+a_2C^{\frac{3}{2}}_2(2u-1)
+a_4C^{\frac{3}{2}}_4(2u-1)\right\}\, , \nonumber\\
\varphi_p(u,\mu)&=&1+\left\{30\eta_3-\frac{5}{2}\rho^2\right\}C_2^{\frac{1}{2}}(2u-1)\nonumber \\
&&+\left\{-3\eta_3\omega_3-\frac{27}{20}\rho^2-\frac{81}{10}\rho^2 a_2\right\}C_4^{\frac{1}{2}}(2u-1)\, ,  \nonumber \\
\varphi_\sigma(u,\mu)&=&6u(1-u)\left\{1
+\left[5\eta_3-\frac{1}{2}\eta_3\omega_3-\frac{7}{20}\rho^2-\frac{3}{5}\rho^2 a_2\right]C_2^{\frac{3}{2}}(2u-1)\right\}\, , \nonumber \\
\phi_{3K}(\alpha_i,\mu) &=& 360 \alpha_u \alpha_s \alpha_g^2 \left
\{1 +\lambda_3(\alpha_u-\alpha_s)+ \omega_3 \frac{1}{2} ( 7 \alpha_g
- 3) \right\} \, , \nonumber\\
V_{\parallel}(\alpha_i,\mu) &=& 120\alpha_u \alpha_s \alpha_g \left(
v_{00}+v_{10}(3\alpha_g-1)\right)\, ,
\nonumber \\
A_{\parallel}(\alpha_i,\mu) &=& 120 \alpha_u \alpha_s \alpha_g
a_{10} (\alpha_s-\alpha_u)\, ,
\nonumber\\
V_{\perp}(\alpha_i,\mu) &=& -30\alpha_g^2
\left\{h_{00}(1-\alpha_g)+h_{01}\left[\alpha_g(1-\alpha_g)-6\alpha_u
\alpha_s\right] \right.  \nonumber\\
&&\left. +h_{10}\left[
\alpha_g(1-\alpha_g)-\frac{3}{2}\left(\alpha_u^2+\alpha_s^2\right)\right]\right\}\,
, \nonumber\\
A_{\perp}(\alpha_i,\mu) &=&  30 \alpha_g^2 (\alpha_u-\alpha_s) \left\{h_{00}+h_{01}\alpha_g+\frac{1}{2}h_{10}(5\alpha_g-3)  \right\}, \nonumber\\
A(u,\mu)&=&6u(1-u)\left\{
\frac{16}{15}+\frac{24}{35}a_2+20\eta_3+\frac{20}{9}\eta_4 \right.
\nonumber \\
&&+\left[
-\frac{1}{15}+\frac{1}{16}-\frac{7}{27}\eta_3\omega_3-\frac{10}{27}\eta_4\right]C^{\frac{3}{2}}_2(2u-1)
\nonumber\\
&&\left.+\left[
-\frac{11}{210}a_2-\frac{4}{135}\eta_3\omega_3\right]C^{\frac{3}{2}}_4(2u-1)\right\}+\left\{
 -\frac{18}{5}a_2+21\eta_4\omega_4\right\} \nonumber\\
 && \left\{2u^3(10-15u+6u^2) \log u+2\bar{u}^3(10-15\bar{u}+6\bar{u}^2) \log \bar{u}
 \right. \nonumber\\
 &&\left. +u\bar{u}(2+13u\bar{u})\right\} \, ,\nonumber\\
 g_K(u,\mu)&=&1+g_2C^{\frac{1}{2}}_2(2u-1)+g_4C^{\frac{1}{2}}_4(2u-1)\, ,\nonumber\\
 B(u,\mu)&=&g_K(u,\mu)-\phi_K(u,\mu)\, ,
\end{eqnarray}
where
\begin{eqnarray}
h_{00}&=&v_{00}=-\frac{\eta_4}{3} \, ,\nonumber\\
a_{10}&=&\frac{21}{8}\eta_4 \omega_4-\frac{9}{20}a_2 \, ,\nonumber\\
v_{10}&=&\frac{21}{8}\eta_4 \omega_4 \, ,\nonumber\\
h_{01}&=&\frac{7}{4}\eta_4\omega_4-\frac{3}{20}a_2 \, ,\nonumber\\
h_{10}&=&\frac{7}{2}\eta_4\omega_4+\frac{3}{20}a_2 \, ,\nonumber\\
g_2&=&1+\frac{18}{7}a_2+60\eta_3+\frac{20}{3}\eta_4 \, ,\nonumber\\
g_4&=&-\frac{9}{28}a_2-6\eta_3\omega_3 \, ,
\end{eqnarray}
 here  $ C_2^{\frac{1}{2}}$, $ C_4^{\frac{1}{2}}$
 and $ C_2^{\frac{3}{2}}$ are Gegenbauer polynomials,
  $\eta_3=\frac{f_{3K}}{f_K}\frac{m_q+m_s}{M_K^2}$ and  $\rho^2={m_s^2\over M_K^2}$
 \cite{LCSR,LCSRreview,Belyaev94,Ball98,Ball06}.
\section*{Acknowledgments}
This  work is supported by National Natural Science Foundation,
Grant Number 10405009,  and Key Program Foundation of NCEPU.

\end{document}